%% file: main.tex
\documentclass[a4paper, oneside, twocolumn, notitlepage, 10pt]{extarticle_ecoc}
\usepackage{ecoc}
\input{package_and_definitions}

\begin{document}
\selectlanguage{english}    %

\title{Optimized Soft-Aided Decoding of OFEC and Staircase Codes}%

\author{
    Lukas Rapp\textsuperscript{(1)}, Sisi Miao\textsuperscript{(2)},
    Laurent Schmalen\textsuperscript{(2)}
}

\maketitle                  %

\begin{strip}
    \begin{author_descr}

        \textsuperscript{(1)} \emph{now with} Research Laboratory of Electronics (RLE), MIT, Cambridge, MA 02139 USA,
   \textcolor{blue}{\uline{rappl@mit.edu}} 

   \textsuperscript{(2)} Communications Engineering Lab (CEL), Karlsruhe Institute of Technology, Karlsruhe, Germany

    \end{author_descr}
\end{strip}

\renewcommand\footnotemark{}
\renewcommand\footnoterule{}

\begin{strip}
    \begin{ecoc_abstract}
    We propose a novel soft-aided hard-decision decoding algorithm for general product-like codes. It achieves error correcting performance similar to that of a soft-decision turbo decoder for staircase and OFEC codes, while maintaining a low complexity. \textcopyright2024 The Author(s)
    \end{ecoc_abstract}
\end{strip}

\section{Introduction}
Next-generation high-throughput optical communication systems require novel high-performance low-complexity FEC schemes. \Acp{PC}~\cite{elias1955coding} and its generalizations, e.g., the zipper code family~\cite{sukmadji2022zipper} and the OFEC code~\cite{openroadmOpenROADMMSA2021}, are suitable candidates. However, traditional \ac{SDD} based on \ac{TPD}~\cite{pyndiah1998near} entails high computational complexity and a high internal decoder data flow. Therefore, even with the existing optimization and simplification methods, e.g.,~\cite{AlDweik2009hybrid,AlDweik2018Ultralight}, \ac{TPD} may not meet the energy efficiency requirements for future optical communication systems. This motivates the research on soft-aided hard-decision decoding algorithms, which use hard-message passing as in \ac{HDD}, but introduce a small amount of soft-information to aid the decoder. Our recent soft-aided decoder based on \acp{DRS}~\cite{miao2022JLT,miao2022Adaptive} achieves a decoding performance close to \ac{TPD} for \acp{PC}. The \ac{DRS} is a reliability measurement of the bits which can be updated with hard-messages. The decoding performance improvement stems from approaching miscorrection-free \ac{EaE} decoding. In this paper, we extend the \ac{DRSD} for general product-like codes. We demonstrate the performance of the decoder on a staircase code~\cite{smithStaircaseCodesFEC2012} and OFEC code~\cite{openroadmOpenROADMMSA2021} with numerical simulations. Significant performance improvements compared to \ac{HDD} are observed.

\section{GPCs and Channel Model}
\Acp{GPC} are extensions of a \ac{PC} where every bit is protected by $d_{\mathrm{v}}\geq 2$ component codes $\mathcal{C}_{\mathrm{c}}$ equipped with a component code decoder $\Dc$. For presentation, we consider all component codes $\mathcal{C}_{\mathrm{c}}$ being the same $[n_{\mathrm{c}},k_{\mathrm{c}}]$ \ac{BCH} code. 
For $d_{\mathrm{v}}=2$, exemplary codes are PCs, staircase codes, zipper codes, and the OFEC code. Codes with $d_{\mathrm{v}}>2$ are investigated recently in e.g.,~\cite{shehadeh2023higher,MiaoISIT24_GPC,xu2024reduced}.
To decode GPCs, an iterative process is executed where a set of component codes are decoded by $\Dc$ at a time using a specific schedule until all the words yield a zero syndrome or the maximum number of decoding iterations is reached. The decoding complexity of GPCs depends on $\Dc$, which is a simple \ac{BDD} in \ac{HDD}, resulting in the ubiquitous \ac{iBDD}. In \ac{TPD}, $\Dc$ is a soft-decision decoder with numerous \ac{BDD} steps followed by soft-message passing.

For \ac{BI-AWGN} channels with ${\sigma_\text{n}^2 =(2 \lEsNO)^{-1}}$ and BPSK modulation, the \ac{PDF} of the received absolute value $r \geq 0$ is \vspace{-1.5ex}\[f_{|R|}(r) = f_{R}\left(r \mid X = +1\right) + f_{R}\left( r \mid X = -1\right),\vspace{-1.5ex}\]
with \(f_R\left(r \mid X = \pm1\right)\) being the PDF of \(\mathcal{N}(\pm 1, \sigma_\text{n}^2)\). The corresponding cumulative distribution function is denoted as \(F_{|R|}(r)\).

\section{Proposed Decoder}
An overview of the proposed DRSD for GPCs is depicted in Fig.~\ref{fig:block_diagram}. At the initialization step, every bit in the received block is assigned a ternary decision in $\{\pm1, ?\}$ and a reliability indicator \ac{DRS} in $\{i_\mathrm{s}, i_\mathrm{s}+1, \dots, i_\mathrm{e}\}$. A bit is classified as an erasure if its magnitude of the channel output is smaller than the threshold $T$. In parallel, the DRS values are initialized by a non-uniform quantizer with thresholds \(t_{i_\mathrm{s}} = 0\) and \(\{ t_{i_\mathrm{s} + 1}, \dots, t_{i_\mathrm{e}}\}\). A bit with received absolute value \(r \geq 0\) is assigned the DRS value \(d \in \{i_{\mathrm{s}}, \dots, i_{\mathrm{e}} - 1\}\) if \(t_{d} \leq r < t_{d+1}\) and \(i_e\) if \(t_{i_e} \leq r\). Here, $T$ and \(\{t_{i_\mathrm{s} + 1}, t_{i_\mathrm{s} + 2}, \dots, t_{i_\mathrm{e}}\}\) are optimizable thresholds. 
During iterative decoding, the component code decoder $\Dc$ decodes every word $\bm{y}\in \{0,1,?\}^{n_{\mathrm{c}}}$ with an \ac{EaED} followed by a \ac{MD} step. 

We first describe the \ac{EaED}. Consider a word $\bm{y}$ containing $E$ erasures. If $E=0$, usual BDD is performed. If $\bm{y}\in \mathcal{C}_{\mathrm{c}}$, the \ac{DRS} of every bit in $\bm{y}$ is increased by one. If $E>0$, the erasures are filled with $J$ pairs of distinct random complementary filling patterns, resulting in at most $2^J$ test patterns. Here, $J=1$ if $E=1$ and $J=\min\{\mathcal{J},E\}$ if $E>1$, where $\mathcal{J}\in\{1,2,3,4\}$ is configured for a performance-complexity trade-off. Then the test patterns are decoded with BDD, yielding a set $\mathcal{I}$ of unique candidate codewords with $|\mathcal{I}|\leq 2^J$. The duplicated candidate codewords are discarded. If $\mathcal{I}=\emptyset$, a decoding failure is declared and we proceed to decode the next word. If $|\mathcal{I}|>0$, let $\{\bm{c}_1,\bm{c}_2,\ldots\bm{c}_{|\mathcal{I}|}\}:=\mathcal{I}$ such that the Hamming distance between $\bm{c}_{i_1}$ and $\bm{y}$ is smaller than the distance between $\bm{c}_{i_2}$ and $\bm{y}$ at the non-erasred position of $\bm{y}$ if $i_1<i_2$. 

Then, we perform the \ac{MD} step for $\bm{c}_{i}\in\mathcal{I}$ using a configurable anchor threshold $T_{\mathrm{a}}$ determined based on the code structure. We start with $i=1$. The bits that have a DRS~$> T_{\mathrm{a}}$ are classified as anchor bits and are not allowed to be flipped by $\Dc$. If such anchor bit flips occur, we proceed to the MD step of the next codeword with $i:=i+1$.
When no such anchor bit flips occur, the output is $\Dc(\bm{y}) = \bm{c}_i$, and the remaining candidate codewords are discarded. In the last step of $\Dc$, the \ac{DRS} of all flipped bits in $\bm{c}_i$ is reduced by one. If all $\bm{c}_i\in \mathcal{I}$ are classified as miscorrections, we reduce the \ac{DRS} of all flipped bits in $\bm{c}_1$ by one. The DRSs are clipped to $[i_\mathrm{s}, i_\mathrm{e}]$.
\vspace{-1ex}
\section{DRSD for Staircase Code}
We evaluate the decoder for an example staircase code based on a $[255,231]$ triple-error-correcting BCH code with $1$-bit shortening. 
A sliding-window decoder as in~\cite{smithStaircaseCodesFEC2012} is used with a sliding window of length $7$ and $8$ decoding iterations. We set $\mathcal{J}=1$ to use a conventional \ac{EaED}. 
For a \ac{BI-AWGN} channel, we let $i_\mathrm{s}=0$, $i_\mathrm{e}=31$ and calculate the thresholds \(\{t_{i_\mathrm{s}}, t_{i_\mathrm{s} + 1}, \dots, t_{i_\mathrm{e}}\}\) so that each DRS value is assigned to the same number of bits:
\vspace{-1.5ex}
\begin{equation}\label{eqn:Thresholds}
    t_{k} = \max\left\{0, F_{|R|}^{-1}\left((k - i_\mathrm{s}) / (i_\mathrm{e} - i_\mathrm{s} + 1)\right)\right\},\vspace{-1.5ex}
\end{equation}
where \(F_{|R|}^{-1}(\cdot)\) is the inverse function of \(F_{|R|}(r)\) and \(t_{i_\text{s}} = 0\).
Let $x\in \{0,1,\ldots,6\}$ be the distance between the current block being decoded and the end of the current decoding window.  We set $T_{\mathrm{a}}=2(x+1)$ for $x<6$ and $T_{\mathrm{a}}=28$ if $x=6$. Fig.~\ref{fig:SCC_soft_aided_decoding} shows that DRSD outperforms the other soft-aided decoders and yields a significant decoding performance gain compared to iBDD.

\begin{figure}
    \centering
    \scalebox{0.65}{\input{plots/block}}
    \caption{Block diagram of the proposed DRSD.}
    \label{fig:block_diagram}
\end{figure}
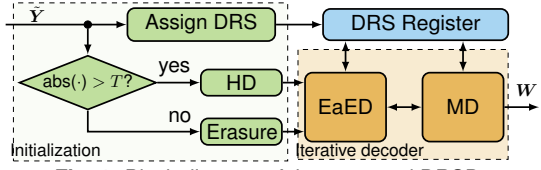

\begin{figure}	
\input{plots/SCC_nu_8_t_3}
 \caption{BER results of a rate 0.811 staircase codes with 1-bit shortened $[255,231],t=3$ BCH code as component code.}
    \label{fig:SCC_soft_aided_decoding}
\end{figure}
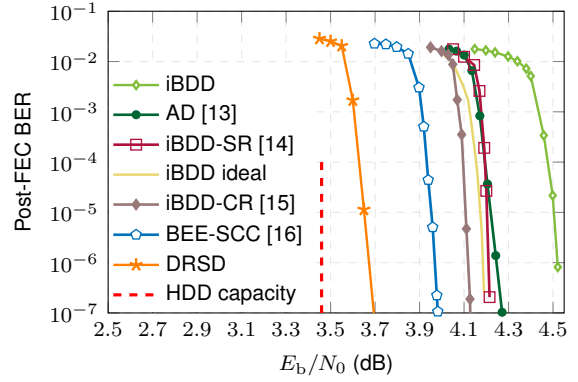

\begin{figure*}
    \centering
    \begin{minipage}{0.36\textwidth}
    \includegraphics{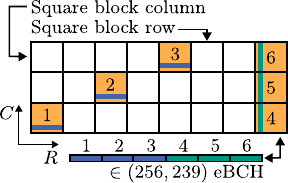}
    \caption{OFEC Code with 3 instead of 8 square block columns for clarity. Blue and green stripes represent the front and back of the component code, respectively.}
    \label{fig:ofec}
    
    \vspace{1ex}

    \begin{center}
        \includegraphics[width=0.3\textwidth]{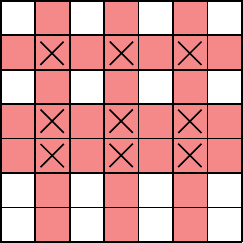}
    \end{center}
    \captionsetup{justification=justified,singlelinecheck=false}
    \vspace{-1em}
    \caption{MSP of a product code with \((t=2)\)-error correcting component codes.    
    }
    \label{fig:stall-patterns}

    \end{minipage}\hfill
    \begin{minipage}{0.61\textwidth}
    \centering
        \input{plots/result_new.tex}
        \vspace{-1em}
    \caption{BER result of the proposed soft-aided DRSD and its variations.}
    \label{fig:results-list}
    \end{minipage}\hfill
\end{figure*}

\section{DRSD for OFEC code}
The OFEC code is a spatially-coupled GPC based on the $[256,239, t=2]$ singly extended BCH code where the code extension allows decoding of one additional erasure. As depicted in Fig.~\ref{fig:ofec}, which is based on Fig.~20 in \cite{openroadmOpenROADMMSA2021}, the code consists of square blocks of \(16 \times 16\) bits organized in a matrix of square block rows and columns. Each component code is split in half (front and back). The back protects a row in a square block row and the front protects multiple columns in diagonally arranged square blocks that were earlier encoded.
Each bit in the structure is part of the front and back of two codewords.
We propose the following modifications to the decoder to ensure compatibility with the structure of the OFEC code.

\emph{DRS Initialization:}
We consider a variation of \eqref{eqn:Thresholds} that only quantizes reliabilities below a threshold 
\(t_{\textnormal{quan,max}} \in \freal_{\geq 0}.
\vspace{-1ex}\):
\begin{equation}\label{eqn:Thresholds_opt}
    t_{k} =
    \max\left\{0,
        F_{|R|}^{-1}\left(\frac{k - i_\mathrm{s}}{i_\mathrm{e} - i_\mathrm{s}} F_{|R|}(t_{\textnormal{quan,max}}) \right)
    \right\}.\vspace{-1ex}
\end{equation}
Bits with received magnitude $|r|>t_{\textnormal{quan,max}}$ are assigned the maximal DRS value \(i_e\). 
This adjustment is made to accommodate some practical implementations, where a maximum quantization value exists for a quantizer.

\emph{Tracking Decoding Iterations:}
For the OFEC code, the anchor threshold is determined as\vspace{-1ex}
\begin{numcases}
    {T_{\textnormal{a}} \;\,\mathclap{=}\;}
    T_{\textnormal{a,step}}
    \left\lfloor 
        \frac{\ell - 1}{p_{\textnormal{a}}} 
    \right\rfloor 
    \,\mathclap{+}\; T_{\textnormal{a,init}}, \quad \mkern-9mu \text{for \(\ell \leq 2 L - p_{\text{r}}\)}, \nonumber \\
    T_{\textnormal{a}}^{*}, \qquad \mkern-18mu
    \hspace{2.57cm}
    \text{for \(\ell > 2 L - p_{\text{r}}\)}.\label{eqn:calc-ta}\vspace{-1em}
\end{numcases}
where \(T_{\textnormal{a,init}}\) is the initial anchor threshold and \(p_\text{a}\), \(p_\text{r}\) and \(T_\text{a,step}\) are new parameters to control \(T_\text{a}\). Therefore, it is essential to know the current decoding iteration \(\ell \in \nb\). 
However, in the OFEC code, new blocks of bits are added in each decoding iteration and there is no single iteration \(\ell\). 
A way to address this issue is to store the number of decoding iterations \(\ell_i\) that have been applied to each bit \(y_i\) and calculate \(T_\text{a}\) individually for each bit using \eqref{eqn:calc-ta}, which requires additional storage. This is used in our simulations.
Alternatively, we derive from the OFEC structure that \(\ell_i\) can be calculated as
\(
\ell_i = 2 \lfloor \widetilde{R}_i / 20 \rfloor + \ind{\mathtt{a}} + \ind{\mathtt{b}}
\)
where condition $\mathtt{a} := \{\lfloor (\widetilde{R}_i \,\%\, 20 - 6) / 2 \rfloor \geq 7 - C_i\}$ and condition $\mathtt{b} := \{\text{\(\widetilde{R}_i\) is the first or the second first buffer row}\}$.
Here, \(R_i\) and \(C_i\) denote the square block row and column index of the bit, respectively, and \(\widetilde{R} = R - R_{\text{current}}\) denotes the relative index with respect to the most recently added row \(R_{\text{current}}\). This eliminates the need for additional storage.

\emph{Stall Pattern Removal (SPR):}
Due to the hard-decision nature of the DRSD, a higher error floor than for TPD is expected. As DRSD effectively avoids miscorrections during iterative decoding, \acp{MSP} are dominant at the error floor region. Figure~\ref{fig:stall-patterns} depicts a \ac{MSP} in a PC block. In the PC, decoding a codeword with \(3\) errors results in a decoding failure or a miscorrection. If these errors are arranged in a \(3 \times 3\) grid, iterative decoding will fail.
As the OFEC code is spatially coupled, errors due to \acp{MSP} are distributed over several square block rows.
However, because every bit is protected by two codewords, they can be understood similarly to \acp{MSP} in a simple PC block.
\Acp{MSP} of the OFEC code can be detected and removed via bit flips \cite{holzbaurImprovedDecodingError2019}:\\
    (1) Two flags are assigned to each bit to indicate whether its last decoding in the front or back of a codeword failed, respectively.\\
    (2) After \(L-1\) full decoding iterations of a square block row, its bits with both flags being positive are flipped, followed by \(1\) decoding iteration to clean up the remaining errors.

\emph{Hyperparameter Optimization:}
DRSD depends on several parameters (\(i_\text{s}\), \(i_\text{e}\), \(T\), \(T_{\text{a,init}}\), \(T_\text{a,step}\), \(T_{\textnormal{a}}^{*}\), \(p_\text{a}\) and \(p_\text{r}\)) that are optimized for the OFEC code with the hyperparameter optimization tool Optuna \cite{optuna_2019}: Optuna iteratively samples the best hyperparameters based on previous simulation results. The objective function is the SNR threshold at which the simulated BER is \num{e-4}. The threshold is simulated using the methods described in \cite{rappErrorandErasureDecodingProduct2022}. In a second optimization round, the parameters are fine-tuned using the BER at a fixed SNR as the objective function.

\section{Simulation Results}
We consider the following transmission model:
 The OFEC encoder (i.e., \emph{ENC0} in Fig.~19 in \cite{openroadmOpenROADMMSA2021}) encodes blocks of \(3552\) uniformly distributed information bits into \(4096\) code bits, which are mapped to QPSK symbols using a Gray code \cite{openroadmOpenROADMMSA2021} and transmitted over an AWGN channel.
We model this as the transmission of two independent  BPSK symbols with an amplitude \(\frac{1}{\sqrt{2}}\) affected by (real-valued) AWGN noise 
\(
  \sigma_\text{n}^2 = (2 \lEsNO)^{-1}.
\)

We evaluate the decoder by simulating the continuous transmission of OFEC codewords until \num{e12} bits are transmitted or \num{e6} bit errors occur.
We compare DRSD with iBDD and the OFEC soft decision decoder (\ac{TPD}).
TPD decodes each OFEC square block row with three soft-decision decoding iterations followed by two hard-decision decoding iterations (Appendix III in \cite{RecommendationITUT7092020}), where the Chase decoder uses \(42\) test patterns.

For $\mathcal{J}=1$, DRSD with \(20\) and \(40\) full decoding iterations outperforms iBDD at a BER of \num{e-9} by around \SI{0.74}{\dB} and \SI{0.78}{\dB} , respectively.
These results use the DRS quantization method \eqref{eqn:Thresholds}. 
Using variation \eqref{eqn:Thresholds_opt} with \(t_{\textnormal{quan,max}} = 0.9\) while keeping the other optimized parameters improves the performance further (labeled as \emph{20 it. var. \eqref{eqn:Thresholds_opt}}).
Increasing $\mathcal{J}$ to 2 and 3 leads to a decoding gain of \SI{0.89}{} and \SI{0.92}{\dB} compared to iBDD, respectively. Further increasing of the $\mathcal{J}$ value does not improve the performance significantly.

An error floor below a BER of \num{e-9} is observed when no SPR is used (dashed curves). However, all error floors can be removed with the aforementioned bit flipping SPR (solid curves), i.e., no bit error occurred when simulating up to $10^{12}$ bits.

DRSD requires more storage and involves ternary message passing compared to the \ac{HDD} iBDD decoder. However, the level of message passing remains significantly lower than that of TPD. The additional BDD steps caused by erasures, as compared to iBDD, are negligible since erasures are resolved after initial iterations, simplifying EaED to BDD.

\section{Conclusion}
We propose a generalized low-complexity hybrid decoding scheme for PC-like codes.
Numerical simulation results for two example codes demonstrate performance improvements compared to traditional HDD.
\section{Acknowledgement}
This work has received funding from the European Research Council (ERC) under the European Union's Horizon 2020 research and innovation programme (grant agreement No. 101001899).
\printbibliography

\vspace{-4mm}

\end{document}

%% file: package_and_definitions.tex
\usepackage{tikz,pgfplots}
\usepackage{graphicx}
\usepackage{placeins}
\usepackage{mfirstuc}
\usepackage{mathtools}
\usepackage{wrapfig}
\usepackage{pdfpages}
\usepackage{diagbox}
\usepackage{makecell}
\usepackage{amssymb}
\usepackage{bbm}
\usepackage{acronym}
\pgfplotsset{compat=1.16}
\usepackage{cases}

\binoppenalty=\maxdimen
\relpenalty=\maxdimen
\usetikzlibrary{backgrounds,shapes,arrows,positioning}
\usetikzlibrary{shapes,snakes}
\usetikzlibrary{fit,calc} %

\usepackage{lipsum}

\usepackage{pgfplots}
\usepackage{siunitx}
\usepackage{xcolor}
\usepackage{tikz}

\newcommand{\freal}{\mathbb{R}}
\newcommand{\nb}{\mathbb{N}}

\newcommand{\lEsNO}{E_{\textnormal{s}} / N_0}

\definecolor{tab0}{RGB}{228,26,28}
\definecolor{tab1}{RGB}{55,126,184}
\definecolor{tab2}{RGB}{77,175,74}
\definecolor{tab3}{RGB}{152,78,163}
\definecolor{tab4}{RGB}{255,127,0}

\usepackage{comment}

\usepackage{mathtools}

\usetikzlibrary{fit}
\usetikzlibrary{decorations.pathreplacing}
\definecolor{KITpalegreen}{RGB}{130,190,60}
\definecolor{KITcyanblue}{RGB}{80,170,230}
\definecolor{KITorange}{rgb}{.87,.60,.10}
\definecolor{tab104}{RGB}{152,78,163}
\definecolor{colorDRSD}{RGB}{255,128,0}
\definecolor{colorideal}{RGB}{224,243,248}
\definecolor{colorDRSDdeter}{RGB}{168,129,188}
\definecolor{colorDRSDplus20}{RGB}{255,102,178}
\definecolor{colorBEE-PC}{RGB}{96,96,96}
\definecolor{colorAD}{RGB}{0,102,51}
\definecolor{orange1}{RGB}{254,196,79}%
\definecolor{orange2}{RGB}{254,153,41}
\definecolor{orange3}{RGB}{217,95,14}
\definecolor{orange4}{RGB}{153,52,4}

\definecolor{pink1}{RGB}{250,159,181}%
\definecolor{pink2}{RGB}{247,104,161}
\definecolor{pink3}{RGB}{197,27,138}
\definecolor{pink4}{RGB}{122,1,119}

\definecolor{mygreen}{rgb}{0.69, 0.87, 0.541}%
\definecolor{myblue}{rgb}{0,0.4470,0.7410}
\definecolor{myblack}{rgb}{0.2,0.2,0.2}
\definecolor{proposed5}{RGB}{188,189,220}
\definecolor{proposed4}{RGB}{158,154,200}
\definecolor{proposed1}{RGB}{128,125,186}
\definecolor{proposed2}{RGB}{106,81,163}
\definecolor{proposed3}{RGB}{84,39,143}
\definecolor{proposedfull}{RGB}{63,0,125}

\definecolor{KITpalegreen}{RGB}{130,190,60}
\definecolor{KITcyanblue}{RGB}{80,170,230}
\definecolor{KITorange}{rgb}{.87,.60,.10}

\definecolor{KITpurple}{RGB}{160,0,120}

\usepackage{bm}

\usepackage{bm}

\newcommand{\ind}[1]{\mathbbm{1}_{\{#1\}}}

\newcommand{\Dc}{\mathsf{D}_{\mathsf{c}}}

\tikzset{mark options={mark size=3, line width=1pt}}
\definecolor{colorCapacityGainAWGN}{RGB}{152,78,163}
\definecolor{myblue}{rgb}{0,0.4470,0.7410}

\acrodef{MC}{miscorrection}
\acrodef{HDD}{hard decision decoding}
\acrodef{HD}{hard decision}
\acrodef{SDD}{soft decision decoding}
\acrodef{TPD}{turbo product decoding}
\acrodef{BSC}{binary symmetric channel}
\acrodef{iBDD}{iterative \ac{BDD}}
\acrodef{BDD}{bounded distance decoding}
\acrodef{SA-HDD}{soft-aided \ac{HDD}}
\acrodef{PC}{product code}
\acrodef{AD}{anchor decoding}
\acrodef{HRB}{highly reliable bit}
\acrodef{EaE}{error-and-erasure}
\acrodef{EaED}{\ac{EaE} decoder}
\acrodef{SABM}{soft-aided bit marking}
\acrodef{iEaED}{iterative error-and-erasure decoding}
\acrodef{BI-AWGN}{binary input additive white Gaussian noise}
\acrodef{BCH}{Bose--Chaudhuri--Hocquenghem}
\acrodef{DRS}{dynamic reliability score}
\acrodef{DRSD}{\ac{DRS} decoder}
\acrodef{BER}{bit error rate}
\acrodef{SABM-SR}{SABM with scaled reliabilities}
\acrodef{NCG}{net coding gain}
\acrodef{GMD}{generalized minimal distance}
\acrodef{GMDD}{generalized minimal distance decoder}
\acrodef{RS}{Reed--Solomon}
\acrodef{DE}{density evolution}
\acrodef{LLR}{log-likelihood ratio}
\acrodef{BPSK}{binary phase shift keying} 
\acrodef{SNR}{signal-to-noise ratio} 
\acrodef{BEE-PC}{binary message passing based on \ac{EaE} decoding for \acp{PC}}
\acrodef{EMP}{extrinsic message passing}
\acrodef{SISO}{soft input soft output}
\acrodef{HIHO}{hard input hard output}
\acrodef{MD}{miscorrection detection}
\acrodef{GPC}{generalized \ac{PC}}
\acrodef{PDF}{probability density function}
\acrodef{MSP}{minimal stall pattern}

%% file: plots/block.tex
\usetikzlibrary{calc} 
\tikzset{block/.style={draw, line width=1pt, text width=4em, minimum height=2em, align=center, rounded corners=1.5mm,fill=white},
line/.style={-latex, line width=1pt}   %
}  
\begin{tikzpicture}[tight background]

    \node[draw=none] (input) {};
    \node[draw=none, right=1em of input, yshift = 0.6em] (tildeY) {$\tilde{\bm{Y}}$};
    
    \node[draw,circle,fill,inner sep=1.5pt, right=4.5em of input] (intersec) {};
      
    \node[block,right=2em of intersec,fill=KITpalegreen!50,text width = 15ex,minimum height = 1em,] (assign) {\makecell{\large Assign  \large DRS}};  
    
    \node [draw, diamond, line width=1pt,  aspect=2.5,fill=KITpalegreen!50,below of =intersec,yshift=-1ex,inner sep=0.1ex](condition){{abs($\cdot$)} {$>T$?}};

    \node[block,fill=KITpalegreen!50,minimum height = 1em,right of =condition,xshift=6em] (hd)  {\large HD}; 

    \node[block,fill=KITpalegreen!50, minimum height = 1em,below of =hd] (erasure)  {\large Erasure}; 
    
    \node[draw, line width=1pt, text width=4em, minimum height=4em, align=center, fill=KITorange!70, rounded corners=1.5mm] (eaed) at ([xshift=6em]$(hd)!0.5!(erasure)$){\large EaED};  
    \node[draw, line width=1pt, text width=4em, minimum height=4em, align=center,right=2em of eaed, fill=KITorange!70, rounded corners=1.5mm] (md) {\large MD};  
    \node[block,draw, line width=1pt, text width=10em, minimum height=1em, align=center,fill=KITcyanblue!50, right of = assign, xshift = 9.5em] (reg) {\large DRS Register};  
    
    \node[draw=none, right=2em of md] (output) {};
    \node[draw=none, right=0.3em of md, yshift = 1em] (W) {$\bm{W}$};

    \node[draw=none] at ([yshift=-2.5em]$(eaed)!0.12!(md)$)(decodertext) {Iterative decoder};
    \node[draw=none, left=10.5em of decodertext] {Initialization};
    
    \begin{scope}[on background layer]     
    \node[draw,inner xsep=1mm,inner ysep=4mm,dashed,line width=0.5pt, fit=(eaed)(md),align=left,yshift=0.3ex,xshift = -0.25mm,fill=KITorange!20](backgroundDecoder){};
    
    \node[draw, fill opacity=0.2,inner xsep=1mm,inner ysep=2mm,dashed,line width=0.5pt,fit=(intersec)(assign)(condition)(hd)(erasure),align=left,yshift=-0.6ex,fill=KITpalegreen!20](backgroundIni){};
    \end{scope}

    \draw[line] (input)-- (assign);  
    \draw[line] (assign) -- (reg); %
    \draw[line] (condition) --(hd) node[above,pos=0.4] {\large yes};
    \draw[line] (condition)  |- (erasure); 
    \node[draw=none, align=left]at ([yshift=-2em]$(condition)!0.6!(hd)$) (no) {\large no};

    \coordinate (midpoint1) at ($(hd.east)$);
    \coordinate (xCoord) at ($(hd.east)$);
    \coordinate (yCoord) at ($(eaed.west)$);
    \coordinate (midpoint2) at (xCoord -| yCoord);
    \draw[-latex, line width=1pt] (midpoint1) -- (midpoint2);

    \coordinate (midpoint11) at ($(erasure.east)$);
    \coordinate (xCoord1) at ($(erasure.east)$);
    \coordinate (yCoord1) at ($(eaed.west)$);
    \coordinate (midpoint22) at (xCoord1 -| yCoord1);
    \draw[-latex, line width=1pt] (midpoint11) -- (midpoint22);

    \draw[line] (intersec) -- (condition);

    \draw[line] ($(md.east)$)-- (output);

     \coordinate (midpoint111) at ($(eaed.north)$);
    \coordinate (xCoord11) at ($(eaed.north)$);
    \coordinate (yCoord22) at ($(reg.south)$);
    \coordinate (midpoint222) at (xCoord11 |- yCoord22);
    \draw[latex-latex, line width=1pt] (midpoint111) -- (midpoint222);

   \coordinate (midpoint111) at ($(md.north)$);
    \coordinate (xCoord11) at ($(md.north)$);
    \coordinate (yCoord22) at ($(reg.south)$);
    \coordinate (midpoint222) at (xCoord11 |- yCoord22);
    \draw[latex-latex, line width=1pt] (midpoint111) -- (midpoint222);

    \draw[latex-latex, line width=1pt] ($(eaed.east)$) -- ($(md.west)$);
\end{tikzpicture}  

%% file: plots/SCC_nu_8_t_3.tex
\definecolor{ColorBEEPC}{rgb}{0,0.4470,0.7410}
\definecolor{ColorAD}{RGB}{0,102,51}
\definecolor{ColorDRSD}{RGB}{255,128,0}
\definecolor{ColoriBDD(ideal)}{rgb}{0.85098, 0.32941, 0.10196}
\definecolor{ColorIGMDDSR}{rgb}{0,0.49804,0}
\definecolor{ColorIBDDSR}{RGB}{176, 18, 62}
\definecolor{ColorSABM}{rgb}{1,0,1}%
\definecolor{ColorSABMSR}{rgb}{1,0,1}%
\definecolor{ColorDRSDlist}{RGB}{84,39,143}
\definecolor{ColoriBDDCR}{rgb}{0.6, 0.47, 0.48}
\definecolor{ColorBMPGMDD}{rgb}{0.75, 0.58, 0.89}
\definecolor{ColoriBDD}{RGB}{130,190,60}
\definecolor{ColoriBDDideal}{rgb}{0.91, 0.84, 0.42}
\begin{tikzpicture}
\pgfplotsset{grid style={gray!20,dashed}}
\pgfplotsset{every tick label/.append style={font=\footnotesize}}
        \begin{axis}[%
        scale only axis=true,
        name=ax1,
        width=6cm,
        height=4cm,
        xmin=2.5,
        xmax=4.55,
        ymode=log,
        ymin=1e-7,
        ymax=0.1,
        yminorticks,
        axis background/.style={fill=white, mark size=1.5pt},
        xmajorgrids,
        ymajorgrids,
        yminorgrids,
        xminorgrids,
        xtick={2.1,2.3,...,5.3},
        ytick={0.1,0.01,0.001,1e-4,1e-5,1e-6,1e-7,1e-8},
        ylabel={Post-FEC BER},
        xlabel={$E_{\mathrm{b}}/N_{\mathrm{0}}$ (dB)},
        label style={font=\footnotesize},
        legend cell align={left},
        legend style={anchor = south west, at={(0.002,0.002)},draw=none, fill opacity=1, text opacity = 1,legend columns=1, row sep = 0pt,font=\footnotesize,inner sep=1pt, outer sep=0pt}
]

\addplot [color=ColoriBDD, line width=0.9pt, mark=diamond*, mark options={solid,fill=white, mark size=1.5pt}]
  table[row sep=crcr]{%
4.147942754919499  0.01771422463351088\\
4.1980322003577815  0.016652387566325352\\
4.23953488372093  0.015177773017322739\\
4.298211091234347  0.012608724076806821\\
4.339713774597495  0.010155736154404196\\
4.379785330948121  0.0072287033509495805\\
4.399821109123435  0.005145284530985864\\
4.458497316636851  0.0003390012833314503\\
4.4985688729874775  0.000021655612406317543\\
4.521466905187835  8.179969377519501e-7\\
} ;\addlegendentry{iBDD}

	\addplot [color=ColorAD, solid, line width=1.0pt, mark=*, mark options={solid, fill=ColorAD,mark size=1.2pt}]
table[row sep=crcr]{%
4.032021466905188  0.01771422463351088\\
4.066368515205725  0.01614558150261839\\
4.100715563506261  0.013412717530678997\\
4.135062611806798  0.006588581861506828\\
4.170840787119857  0.0008307361074919377\\
4.205187835420394  0.00003662337713903384\\
4.242397137745975  0.0000013833739627296252\\
4.2710196779964225  1.0313897683787221e-7\\
 };
 \addlegendentry{AD~\cite{hager2018approaching}}

 \addplot [color=ColorIBDDSR, solid, line width=1.0pt, mark=square, mark options={solid, line width = 0.5pt, fill=ColorIBDDSR}]
table[row sep=crcr]{%
4.050666666666667 0.01759069582234984\\
4.1000000000000005 0.012426236719115056\\
4.145333333333333 0.008527419141916296\\
4.169333333333333 0.0026007244442001164\\
4.190666666666667 0.00019187520990478384\\
4.2 0.000026771515454209492\\
4.214666666666667 2.0628393573248143e-7\\
};
\addlegendentry{\footnotesize{iBDD-SR~\cite{sheikh2019binary}}}

     \addplot [color=ColoriBDDideal, line width=0.9pt, mark= circle, mark options={solid,  mark size=1.5pt}]    table[row sep=crcr]{%
3.9976744186046513  0.014715846019280589\\
4.047763864042934  0.009846652027941251\\
4.097853309481216  0.0031379375436648305\\
4.119320214669052  0.0016397026580002088\\
4.139355992844365  0.000349642455095314\\
4.159391771019678  0.00006388062072657218\\
4.179427549194991  0.000004208826990208906\\
4.189445438282648  2.6067980205769284e-7\\
 };
\addlegendentry{iBDD ideal}

	\addplot [color=ColoriBDDCR, solid, line width=1.0pt, mark=diamond*, mark options={solid, line width = 0.5pt, fill=ColoriBDDCR}]
table[row sep=crcr]{%
3.949333333333334  0.019187520990478345\\
3.9986666666666673  0.016126761746240677\\
4.029333333333334  0.013554249376408268\\
4.049333333333334  0.00877801313597249\\
4.069333333333334  0.0017337788895633694\\
4.089333333333334  0.00035250867676009205\\
4.110666666666667 0.000004709295850583372\\
4.126666666666667 1.8911656009694758e-7\\
4.128 9.437191487625724e-8\\
};
\addlegendentry{\footnotesize{iBDD-CR~\cite{sheikh2021refined}}}

\addplot [color=ColorBEEPC, solid, line width=1.0pt, mark=pentagon*, mark options={solid, line width = 0.5pt, ColorBEEPC, fill=white}]
  table[row sep=crcr]{%
3.6971377459749553  0.02268321963696042\\
3.7486583184257602  0.021992868586060324\\
3.7973166368515208  0.019435271149298293\\
3.8488372093023258  0.01426797751000861\\
3.897495527728086  0.0030424361767690065\\
3.9189624329159214  0.000506638263613666\\
3.9389982110912345  0.0000440854524183413\\
3.9590339892665476  0.00000506638263613666\\
3.9776386404293382  2.2335377063851325e-7\\
3.981932021466905  1.0637648543163142e-7\\
};
\addlegendentry{BEE-SCC~\cite{sheikh2021novel}}

  \addplot [color=ColorDRSD, line width=0.9pt, mark=star, mark options={solid, ColorDRSD, mark size=2.5pt}]
   table[x=EbNo, y=BER, col sep=semicolon,row sep=crcr]{%
EbNo;EsNo;delta;decodedFrame;FrameErr;FER;BER\\
3.45;2.54034;0.0290686;177;177;1;0.0284729\\
3.5;2.59034;0.0283511;177;177;1;0.025202\\
3.55;2.64034;0.0276443;177;177;1;0.0205311\\
3.6;2.69034;0.0269484;196;159;0.811224;0.00169746\\
3.65;2.74034;0.0262632;379;81;0.21372;1.13815e-05\\
3.7;2.79034;0.0255888;1484;54;0.0363881;5.60219e-08\\
 };
\addlegendentry{DRSD}
\addplot [color=red, dashed, line width = 1.2pt]
  table[row sep=crcr]{%
3.4595 0.0001\\
3.4595 0.000000000001\\
};
\addlegendentry{HDD capacity}

\end{axis}

\end{tikzpicture}  

%% file: plots/result_new.tex
        \begin{tikzpicture}
\pgfplotsset{grid style={gray!20,dashed,line width = 0.3pt},every tick label/.append style={font=\footnotesize},
legend image code/.code={
\draw[mark repeat=2,mark phase=2]
plot coordinates {
(0cm,0cm)
(0.15cm,0cm)        %
(0.3cm,0cm)         %
};%
}
}

    \begin{axis}[
    scale only axis=true,
        name = ax1,
        compat=newest,
        ymode=log,
        xlabel= \(\lEsNO\) (dB),
        ylabel= Post-FEC BER,
        xmin=6.1, xmax=7.6,
        xtick={6.1,6.2,...,7.6},
        ytick = {0.001,0.0001,0.00001,0.000001,0.0000001,0.00000001,0.000000001,0.0000000001,0.00000000001,0.000000000001},
        ymin = 1e-11, ymax = 1e-3,
        width=8cm,
        height=5.5cm,
        grid=major,
        legend pos=south west,
        legend cell align={left},
        legend style={
            font=\scriptsize,
            at={(axis cs:6.8,2e-8)},anchor=south west
        },
        legend columns=1, 
        xmajorgrids,
        ymajorgrids,
        yminorgrids,
        xminorgrids,
        xminorticks=true,
        yminorticks=true,
        legend cell align={left},
    ]
    
\addplot[thick,tab2, mark=x, mark options={solid, scale=0.8}]    table[row sep=crcr]{%
6.5 1e-12\\
};
\addlegendentry{20 it.}

\addplot[thick,tab1, mark=o, mark options={solid, scale=0.8}]    table[row sep=crcr]{%
6.5 1e-12\\
};
\addlegendentry{40 it.}

\addplot[thick,tab4, mark=triangle, mark options={solid, scale=0.8}]    table[row sep=crcr]{%
6.5 1e-12\\
};
\addlegendentry{20 it. var. \eqref{eqn:Thresholds_opt}}

\addplot[thick, magenta, mark=square, mark options={solid, scale=0.8}] table[row sep=crcr]{%
6.5 1e-12\\
};
\addlegendentry{20 it.,$\mathcal{J}\;\,\mathclap{=}\;\,2$}

\addplot[thick, KITpurple, mark=diamond, mark options={scale=0.8}] table[row sep=crcr]{%
6.5 1e-12\\
};
\addlegendentry{20 it.,$\mathcal{J}\;\,\mathclap{=}\;\,3$}

\addplot[thick,white, dashed, mark=x, mark options={solid, scale=0.8}]    table[row sep=crcr]{%
6.5 1e-12\\
};
\addlegendentry{}

\addplot[thick,black, dashed, mark=x, mark =none]    table[row sep=crcr]{%
6.5 1e-12\\
};
\addlegendentry{w/o SPR}

\addplot[thick,black, mark=x, mark =none]    table[row sep=crcr]{%
6.5 1e-12\\
};
\addlegendentry{w/ SPR}

\addplot[thick,tab2, dashed, mark=x, mark options={solid, scale=0.8}] table[col sep=semicolon, skip first n=36, x index=0, y expr={\thisrowno{1} < 1e-12 ? 9e-12: \thisrowno{1}}] {plots/31_07_stall_pattern_removal_off_of_17_07_ber_curve_ber_sb_1e9_i20_b5_all_params_trial2.txt};

\addplot[thick,tab4, dashed, mark=triangle, mark options={solid, scale=0.8}] table[col sep=semicolon, skip first n=31, x index=0, y expr={\thisrowno{1} < 1e-12 ? 1e-12: \thisrowno{1}}] {plots/19_07_17_07_ber_curve_maximal_quantized_received_value_0.9_ber_curve_hq.txt};

\addplot[thick,tab4, mark=triangle, mark options={solid, scale=0.8}] table[col sep=semicolon, skip first n=0
, x index=0, y expr={\thisrowno{1} < 1e-12 ? 1e-12: \thisrowno{1}}] {plots/list_1_with_SPR.txt};

\addplot[thick,tab2, mark=x, mark options={solid, scale=0.8}] table[col sep=semicolon, skip first n=36, x index=0, y expr={\thisrowno{1} < 1e-12 ? 9e-12: \thisrowno{1}}] {plots/31_07_stall_pattern_removal_of_17_07_ber_curve_ber_sb_1e9_i20_b5_all_params_trial2.txt};

\addplot[thick,tab1, dashed, mark=o, mark options={solid, scale=0.8}] table[col sep=semicolon, skip first n=26, x index=0, y expr={\thisrowno{1} < 1e-12 ? 9e-12: \thisrowno{1}}] {plots/17_07_ber_curve_tb_1e-4_i40_b5_11_07_trial1_no_increase_step.txt};
    
\addplot[thick,tab1, mark=o, mark options={solid, scale=0.8}] table[col sep=semicolon, skip first n=36, x index=0, y expr={\thisrowno{1} < 1e-12 ? 9e-12: \thisrowno{1}}] {plots/31_07_stall_pattern_removal_of_17_07_ber_curve_tb_1e-4_i40_b5_11_07_trial1_no_increase_step.txt};

\addplot[thick, tab0, mark=x, dashed, mark options={solid, scale=1}] table[col sep=semicolon, skip first n=14, x index=0, y expr={\thisrowno{1} < 1e-12 ? 1e-12: \thisrowno{1}}] {plots/17_07_reference_model_ber_curve.txt}node [pos=0.6,anchor=south,font=\footnotesize,sloped] {TPD};

    \addplot[thick, magenta, dashed, mark=square, mark options={solid, scale=0.8}] table[col sep=semicolon] {plots/list_20_2EaED.txt};

        \addplot[thick, magenta, mark=square, mark options={solid, scale=0.8}] table[col sep=semicolon] {plots/list_2_with_SPR.txt};

\addplot[thick, KITpurple, dashed, mark=diamond, mark options={scale=0.8}] table[col sep=semicolon] {plots/list_20_3EaED.txt};

\addplot[thick, KITpurple, mark=diamond, mark options={scale=0.8}] table[col sep=semicolon] {plots/list_3_with_SPR.txt};

\addplot[thick, black, dashed, mark=x, mark options={solid, scale=1}] table[skip first n=7, col sep=semicolon, x index=0, y expr={\thisrowno{1} < 1e-12 ? 1e-12: \thisrowno{1}}] {plots/iBDD_curve_20_iter_QPSK.txt}node [pos=0.5,anchor=south,font=\footnotesize,sloped] {iBDD};

\addplot[thick, black, mark=x, mark options={solid, scale=1}] table[skip first n=12, col sep=semicolon, x index=0, y expr={\thisrowno{1} < 1e-12 ? 1e-12: \thisrowno{1}}] {plots/iBDD_remove_stall_pattern_run_2.txt};

\draw[thick,latex-latex] (axis cs:6.47,1e-9)--(axis cs:7.39, 1e-9);
\node[font=\footnotesize] at (axis cs:6.93,5e-10) {\SI{0.92}{\dB}};

\draw[thick,latex-latex] (axis cs:6.47,1e-9)--(axis cs:6.21, 1e-9);
\node[font=\footnotesize] at (axis cs:6.34,5e-10) {\SI{0.26}{\dB}};

\end{axis}
\end{tikzpicture}